\documentstyle[12pt]{article}

\newcommand{\be}{\begin{equation}}
\newcommand{\ee}{\end{equation}}
\newcommand{\bea}{\begin{eqnarray}}
\newcommand{\eea}{\end{eqnarray}}
\newcommand{\nn}{\nonumber\\}

\def\L{{\cal X}}                       
\def\Q{{X}}                            
\def\J{{\cal J}}                       
\def\C{{J}}                            
\def\X{{\xi}}                          
\def\Y{{\eta}}                         
\def\T{{\theta}}                       
\def\R{{\mathbf R}}                    
\def\A{{\cal A}}                       
\def\B{{\cal B}}                       
\def\E{{\cal E}}                       
\def\F{{\cal F}}                       
\def\G{{\cal G}}                       
\def\M{{\cal M}}                       
\def\O{{\cal O}}                       

\begin{document}

\renewcommand{\thefootnote}{\fnsymbol{footnote}}
\fnsymbol{footnote}

\vspace{1cm}

\begin{center} 
\Large \bf Adler-Kostant-Symes systems as Lagrangian gauge theories\\
\end{center}

\vspace{.1in}

\begin{center}
L.~Feh\'er\footnote{Corresponding author,  e-mail: lfeher@sol.cc.u-szeged.hu}
and A. G\'abor\\

\vspace{0.2in}

Department of  Theoretical Physics,  University of Szeged\\
Tisza Lajos krt.~84-86, H-6720 Szeged, Hungary

\vspace{.3in}

{\bf  Abstract}\\

\end{center}

\vspace{.1in}

{\parindent=25pt
\narrower\small

It is well known that the integrable Hamiltonian systems defined by 
the Adler-Kostant-Symes construction correspond via Hamiltonian reduction
to systems on cotangent bundles of Lie groups.
Generalizing previous results on Toda systems, 
here a Lagrangian version of the reduction 
procedure  is exhibited for those cases for which the 
underlying Lie algebra admits an invariant scalar product.
This is achieved by constructing a Lagrangian with gauge symmetry  
in such a way that, by means of the Dirac algorithm, this
Lagrangian reproduces the Adler-Kostant-Symes system whose 
Hamiltonian is the quadratic form associated with the scalar 
product on the Lie algebra.
}

\newpage

\section{Introduction}
\setcounter{equation}{0}

\renewcommand{\thefootnote}{\arabic{footnote}}
\setcounter{footnote}{0}

The Adler-Kostant-Symes (AKS) construction  associates
Hamiltonian systems that are in  many cases integrable 
with certain Lie algebraic data \cite{Adler, Kostant, Symes}. 
As found by Reyman and Semenov-Tian-Shansky \cite{RSTS}, 
these systems may be viewed as  symmetry reductions of 
corresponding Hamiltonian systems on cotangent bundles of Lie groups generated 
by Hamiltonians invariant under left and right translations.
An advantage of such a viewpoint is that 
it leads to a natural regularization of some 
AKS systems whose Hamiltonian vector field is 
incomplete \cite{RSTS, Reyman, RSTSreview, FT}. 

The aim of this paper is to provide a Lagrangian 
description for an important subclass of the AKS systems.
Our construction requires the underlying Lie algebra to be self-dual
and a further technical condition must hold.
These conditions are satisfied, for example, in the case of the
open Toda lattices and their generalizations that are among 
the most studied integrable systems.
The conformal Toda field theories were treated in a similar Lagrangian
manner in \cite{Balog}, which actually served as the starting point for 
the present work.  
Our Lagrangian may be used in the future to perform a path integral quantization
of the  AKS systems, and it may permit interesting generalizations 
in the field theoretical case in analogy with the Toda systems.

Let $G$ be a connected real Lie group whose Lie algebra $\G$ 
is equipped\footnote{For the structure of such `self-dual' 
Lie algebras, see e.g.~\cite{Figu}.}  
with a nondegenerate, symmetric, $G$-invariant 
bilinear form $\langle\ ,\ \rangle$.
Identify $\G^*$ with $\G$ by means of 
the `scalar product' $\langle\ ,\ \rangle$.  
Suppose that $\A,\B\subset \G$ are Lie subalgebras in such a way
that as a vector space
\be
\G=\A + \B.
\label{1.1}\ee
This induces the decomposition
\be
\G=\A^\perp + \B^\perp,
\label{1.2}\ee  
which gives rise to the further identifications $\A^* \cong \B^\perp$ and 
$\B^*\cong \A^\perp$.
We denote by $\pi_\A$, $\pi_\B$ and by $\pi_{\A^\perp}$, $\pi_{\B^\perp}$
the projection operators on $\G$ associated with these decompositions.
Let $A,B\subset G$ be the connected Lie subgroups corresponding to $\A,\B$
and fix elements $\mu\in \A^*$ and $\nu\in \B^*$.
The phase space of the AKS system of our interest, 
designated as $\M_{\mu, \nu}$, 
consists of those elements $\L\in \G$ that have the following form: 
\be
\L=\L_{\A^*} + \L_{\B^*}
\quad\hbox{with}\quad
\L_{\A^*}\in \O^-_A(\mu),\quad
\L_{\B^*}\in \O^+_B(\nu),
\label{1.3}\ee
where 
\be
\O^-_A(\mu)= \{ \pi_{\B^\perp}( g \mu g^{-1}) \,\vert\, \forall g\in A\},\qquad
\O^+_B(\nu)= \{ \pi_{\A^\perp}( g \nu g^{-1}) \,\vert\, \forall g\in B\}
\label{1.4}\ee
are the coadjoint orbits of $A$ and $B$ through $\mu$ and $\nu$, respectively.   
The plus/minus superscripts indicate that these orbits are equipped with opposite
Lie-Poisson brackets. 
In the AKS construction the Poisson brackets, denoted here by $\{\ ,\ \}_*$,  
are postulated to be
\bea
&&\{\langle \L_{\A^*}, \xi\rangle , \langle \L_{\A^*}, \xi'\rangle\}_*=
-\langle \L_{\A^*}, [\xi ,  \xi']\rangle,
\qquad\forall \xi,\xi'\in \A,\nonumber\\
&&\{\langle \L_{\B^*}, \eta\rangle , \langle \L_{\B^*}, \eta'\rangle\}_*=
\langle \L_{\B^*}, [\eta,  \eta']\rangle,
\qquad\forall \eta, \eta'\in \B,\nonumber\\
&&\{\langle \L_{\A^*}, \xi\rangle , \langle \L_{\B^*}, \eta\rangle\}_*=0.
\label{1.5}\eea  
The main point is that the $G$-invariant functions on $\G^*$ yield a commuting
family with respect to $\{\ ,\ \}_*$  and  generate Hamiltonian systems 
on $\M_{\mu, \nu}$ that are often 
integrable in the Liouville 
sense \cite{Adler, Kostant, Symes, RSTS, Reyman, RSTSreview}. 
In our case a distinguished $G$-invariant 
Hamiltonian is furnished by
\be
H(\L):= \frac{1}{2} \langle \L, \L\rangle.
\label{1.6}\ee     
The evolution equation associated with 
the Hamiltonian system $(\M_{\mu, \nu}, \{\ ,\ \}_*, H)$ reads as 
\be
\dot{\L} = \{ \L, H\}_* = - [ \pi_\A(\L), \L]= [\pi_\B(\L), \L].
\label{1.7}\ee

In this paper we present a Lagrangian model of 
the system given by $(\M_{\mu, \nu}, \{\ ,\ \}_*, H)$. 
The equivalence of the Lagrangian and Hamiltonian descriptions 
is established 
at the level of the equations of motion in section 2.
Then the Poisson bracket aspect is dealt with by applying 
the Dirac algorithm \cite{Dirac, Sund} to the Lagrangian in section 3.
Examples are contained in section 4.
In addition to the above-mentioned data, our construction 
relies on the existence of 
an open submanifold  $\check G\subset G$ which is  
diffeomorphic to $A\times B$ by the map $A\times B \ni(g_A, g_B)\mapsto g_A g_B$.
A typical example, related to Toda type systems,
for which this condition is satisfied is 
$\A=\G_{>0}$ and $\B=\G_{\leq 0}$ for some integral gradation 
$\G=\oplus_{n\in \mathbf Z} \G_n$ 
of a semisimple Lie algebra $\G$.

A remark is in order here concerning our notations. 
Throughout the paper, we pretend that $G$ is a matrix group to simplify notations. 
This is not a real restriction in any sense since one can
rewrite all equations in a more general notation. 
For instance,  $\pi_{\B^\perp}( g \mu g^{-1})$ in (\ref{1.4}) would then 
be replaced by $({\mathrm Ad}_A^* g)(\mu)$ to denote the coadjoint action 
of $g\in A$ on $\mu\in \A^*$ and so on.

\section{The AKS system as a gauge theory}
\setcounter{equation}{0}

Motivated by the work on Toda theories \cite{Balog}, we propose to consider 
the following Lagrangian:
\bea
&&L(g,\dot{g}, \alpha, \beta):= 
\frac{1}{2}\langle \dot{g} g^{-1}, \dot{g} g^{-1} \rangle 
+  \langle \alpha, \dot{g} g^{-1}-\mu\rangle  
+\langle \beta, g^{-1}\dot{g} -\nu\rangle  
 \nonumber\\
&&\phantom{ L(g,\dot{g}, \alpha, \beta):=}
+\langle \alpha, g\beta g^{-1}\rangle +\frac{1}{2}\langle \alpha, \alpha\rangle 
+\frac{1}{2} \langle \beta, \beta\rangle.   
\label{2.1}\eea
Here $g\in G$, $\dot{g}\in T_g G$  and $\alpha\in \A$, $\beta\in \B$. 
The first term is the Lagrangian of a free particle moving on the group
manifold $G$.
The variables $\alpha$ and $\beta$ act essentially as  
Lagrange multipliers that impose the constraints that appear in 
the Hamiltonian reduction treatment \cite{RSTS,Reyman,RSTSreview} of the AKS system.
The terms in the second line are chosen so as to equip the Lagrangian with 
the gauge symmetry that we describe next.

The little groups of the constants $\mu\in \A^*$ and $\nu\in \B^*$ are given by 
\be
A_\mu= \{ a\in A\vert \pi_{\B^\perp}( a \mu a^{-1}) =\mu\},
\qquad
B_\nu= \{ b\in B\vert \pi_{\A^\perp}( b \nu b^{-1}) =\nu\}.
\label{2.2}\ee
We associate a gauge transformation with any curve 
$a(t)\in A_\mu$, $b(t)\in B_\nu$ by letting any curve 
$(g(t),\alpha(t),\beta(t))$ in the configuration space of our Lagrangian 
system transform as  
\bea
&& g(t) \mapsto a(t) g(t) b^{-1}(t) \nonumber\\
&& \alpha(t) \mapsto a(t) \alpha(t)a^{-1}(t) - \dot{a}(t) a^{-1}(t)\nonumber\\
&& \beta(t) \mapsto b(t) \beta(t) b^{-1}(t) + \dot{b}(t) b^{-1}(t). 
\label{2.3}\eea
One can check that $L$ changes by a total time derivative
under these transformations for any $(a(t), b(t))\in A_\mu\times B_\nu$, 
and $A_\mu \times B_\nu \subset A\times B$ is the 
maximal subgroup with this property.

For the further analysis it is 
convenient to introduce the quantities $\C^l$ and $\C^r$ by
\be
\C^r:= \dot{g} g^{-1} + g\beta g^{-1}+ \alpha, 
\qquad
\C^l:= g^{-1} \C^r g = g^{-1}\dot{g} + g^{-1} \alpha g + \beta.
\label{2.4}\ee
Under (\ref{2.3}) their gauge transformation properties are  
\be
\C^r(t)\mapsto a(t) \C^r(t) a^{-1}(t),
\qquad
\C^l(t) \mapsto b(t) \C^l(t) b^{-1}(t). 
\label{2.5}\ee
The Euler-Lagrange equations of $L$ obtained by varying $\alpha$ and 
$\beta$, respectively, are  
\be
\pi_{\B^\perp}(\C^r) = \mu
\quad\hbox{and}\quad
\pi_{\A^\perp} (\C^l)= \nu.
\label{2.6}\ee
The equations that result by varying $g$ are encoded 
by either of the following two relations:
\be
\dot{\C}^r= [ \C^r, \alpha]
\quad\hbox{and}\quad
\dot{\C}^l= [ \beta, \C^l],
\label{2.7}\ee
which are actually equivalent among each other.
It can be verified that the gauge transformations (\ref{2.3}) map any
solution of (\ref{2.6}), (\ref{2.7}) into another solution. 

We remark that the derivation of (\ref{2.7}) is very easy in the case 
for which $\G=gl_n$ and $\langle X,Y\rangle = {\mathrm tr}(XY)$, since in this 
case one can parametrize $g\in GL_n$ by its matrix elements.
In general one derives the Euler-Lagrange equations by using some
arbitrary local coordinates on $G$, and then rewrites those equations
in the coordinate independent form (\ref{2.7}).   

By assumption, there exists an open submanifold $\check G \subset G$ 
diffeomorphic to $A\times B$ by the factorization map
\be 
A\times B \ni (g_A, g_B)\mapsto g_Ag_B\in \check G.
\label{2.8}\ee
 {}From now on we restrict\footnote{It may happen that $\check G=G$, 
examples are mentioned in section 4.}  
$g$ to belong to $\check G$.
By using the decomposition $g=g_A g_B$, 
the first line of the transformation rule (\ref{2.3}) becomes 
\be
g_A(t) \mapsto a(t) g_A(t),
\qquad
g_B(t) \mapsto g_B(t) b^{-1}(t).
\label{2.9}\ee    
If $g\in \check G$, we can define the gauge invariant quantity
\be
\Q:= g_A^{-1} \C^r g_A = g_B \C^l g_B^{-1}.
\label{2.10}\ee
We next show that $\Q$ satisfies the evolution 
equation (\ref{1.7}) of the AKS system.

First, notice that by using the Euler-Lagrange 
equations (\ref{2.6}) $\Q$  can be  
written as 
\be
\Q= \pi_{\B^\perp} ( g_A^{-1} \mu g_A) + \pi_{\A^\perp}(g_B \nu g_B^{-1}).
\label{2.11}\ee 
This follows from 
$\pi_{\B^\perp}(\Q)= \pi_{\B^\perp}( g_A^{-1} \C^r g_A)$ 
by inserting that $\C^r = \mu + \pi_{\A^\perp}(\C^r)$ where 
the second term does not contribute 
since $g_A^{-1} \A^\perp g_A \subset \A^\perp$;
$\pi_{\A^\perp}(\Q)$ is determined similarly.
Upon comparison with (\ref{1.3}), we see that $\Q(t)$
belongs to the AKS phase space $\M_{\mu,\nu}$.  
Second, let us show that (\ref{2.7}) implies  
\be
\dot{\Q} = - [ \pi_\A(\Q), \Q].
\label{2.12}\ee
For this note from (\ref{2.10}) that 
\be
\pi_\A(\Q)= g_A^{-1} \dot{g}_A + g_A^{-1} \alpha g_A,
\qquad
\pi_\B(\Q)= \dot{g}_B g_B^{-1} + g_B \beta g_B^{-1}.
\label{2.13}\ee 
By using the first equation in (\ref{2.7})  we obtain
\be
\dot{\Q} = \frac{d}{dt} ( g_A^{-1} \C^r g_A) =
g_A^{-1} \dot{\C}^r  g_A - [ g_A^{-1} \dot{g}_A, \Q] =
- [g_A^{-1}  \alpha g_A, \Q]   - [ g_A^{-1} \dot{g}_A, \Q],
\label{2.14}\ee 
which gives (\ref{2.12}) on account of (\ref{2.13}).
A similar calculation using $\Q=g_B \C^l g_B^{-1}$ and the second relation 
in (\ref{2.7}) yields $\dot{\Q}= [ \pi_\B(\Q), \Q]$, which is 
plainly equivalent to (\ref{2.12}).  

In conclusion, we have shown that if $g(t)\in \check G$ and 
$(g(t), \alpha(t), \beta(t))$ satisfies the Euler-Lagrange equations of
$L$ in (\ref{2.1}), then the gauge invariant 
function $\Q(t)$ belongs to $\M_{\mu,\nu}$ and satisfies the 
same evolution equation as defined by the AKS system 
$( \M_{\mu,\nu}, \{\ ,\ \}_*, H)$.
Next we explain that the Lagrangian $L$ encodes 
the Hamiltonian structure of the system as well.

\section{Dirac analysis of the Lagrangian}
\setcounter{equation}{0}

The Lagrangian $L$ (\ref{2.1}) is singular since it does 
not depend on the velocities of $\alpha$ and $\beta$.
Thus one has  to apply the Dirac algorithm \cite{Dirac, Sund} to associate
a Hamiltonian system with $L$.
In this manner we below recover the AKS system.
  
The phase space corresponding to the configuration space 
$G \times \A \times \B$ of our Lagrangian system is the
cotangent bundle $\M:= T^* G \times T^*\A \times T^* \B$.
By identifying $T^* G$ with $G\times \G^*$ with the aid of
right translations on $G$ and using as earlier that $\G^*\cong \G$, we have
\be
\M= G \times \G \times \A\times \A^* \times \B \times \B^* =
\{ (g, \J^r, \alpha, \pi_\alpha, \beta, \pi_\beta)\}.
\label{3.1}\ee
Let $\{ \T_a\}$ denote a basis of $\G$ with dual basis $\{ \T^a\}$.
$\{ \T_a\}$ can be chosen as $\{ \T_a\} = \{ \X_m\} \cup 
\{ \Y_r\}$, where $\{ \X_m\}$ and $\{ \Y_r\}$ are bases of
$\A$ and $\B$, respectively. Then
$\{ \T^a\} = \{ \X^m\} \cup \{ \Y^r\}$, where $\{ \X^m\}$ is a basis
of $\A^*\cong\B^\perp$ and $\{ \Y^r \}$ is a basis of $\B^*\cong \A^\perp$. 
Now the fundamental Poisson brackets on $\M$ are given by
\bea
&& \{g, \langle \J^r, \T_a  \rangle\}= \T_a g \nn
&& \{ \langle \J^r, \T_a \rangle, \langle \J^r, \T_b \rangle \}
= \langle \J^r, [\T_a,\T_b] \rangle \nn
&&\{ \langle \alpha, \X^m \rangle , 
\langle \pi_\alpha, \X_n \rangle \} = \delta^m_n\nn
&&\{ \langle \beta, \Y^r \rangle , 
\langle \pi_\beta, \Y_s\rangle \} = \delta^r_s.
\label{3.2}\eea 
The other Poisson brackets 
between $g$, $\J^r$, $\alpha$, $\pi_\alpha$, 
$\beta$ and $\pi_\beta$ vanish. 
We introduce    
\be
\J^l:= g^{-1} \J^r g,
\label{3.3}\ee
and note that it has the Poisson brackets  
\bea
&& \{g, \langle \J^l, \T_a \rangle\}=  g \T_a\nn
&& \{ \langle \J^l, \T_a \rangle, \langle \J^l, \T_b \rangle \}
= - \langle \J^l, [\T_a,\T_b]   \rangle \nn
&& \{ \langle \J^r, \T_a \rangle, \langle \J^l, \T_b  \rangle \}=0.
\label{3.4}\eea
If $q^i$ denotes local coordinates on some $U\subset G$
and $q^i, p_j$ are the corresponding canonical coordinates  on 
$T^* U \subset T^*G$, then on $T^*U$ we have 
\be
\J^r(q,p)=\E^{-1}(q)_a^{\phantom{a}i} p_i \T^a 
\quad\hbox{and}\quad
\J^l(q,p)=\F^{-1}(q)_a^{\phantom{a}i} p_i \T^a ,
\label{3.5}\ee
where $\E^{-1}$ and $\F^{-1}$ are the inverse matrices to $\E$ and $\F$
defined by 
\be
\frac{\partial g(q)}{\partial q^i} g^{-1}(q)= \E(q)_i^{\phantom{i}a} \T_a
\quad\hbox{and}\quad
g^{-1}(q)\frac{\partial g(q)}{\partial q^i} = 
\F(q)_i^{\phantom{i}a} \T_a.
\label{3.6}\ee
The local Poisson brackets $\{ q^i, p_j\}=\delta^i_j$ on $T^*U$
are equivalent to the Poisson brackets of 
$g$, $\J^r$ and $\J^l$ in (\ref{3.2}), (\ref{3.4}). 

Later we shall restrict ourselves to the open submanifold 
$\check \M = T^* \check G \times T^* \A \times T^* \B \subset \M$,
where the factorization $g=g_A g_B$ is valid (\ref{2.8}).
We use also the decompositions
\be
\J^r=\J^r_{\A^*} + \J^r_{ \B^*},
\qquad
\J^l=\J^l_{\A^*} + \J^l_{ \B^*},
\label{3.7}\ee
where $\J^r_{\A^*}= \pi_{\B^\perp}(\J^r)$, 
$\J^r_{\B^*}= \pi_{\A^\perp}(\J^r)$ and similarly for $\J^l$.
On $\check \M$ we thus obtain,
\bea
&& \{g_A, \langle  \J^r_{\A^*}, \X_m \rangle\}= \X_m g_A \nn
&& \{g_B, \langle  \J^r_{\A^*}, \X_m\rangle\}= 0\nn
&& \{g_B, \langle  \J^l_{\B^*}, \Y_r \rangle\}= g_B \Y_r \nn
&& \{g_A, \langle  \J^l_{\B^*}, \Y_r \rangle\}= 0.
\label{3.8}\eea

Now we apply the Dirac algorithm to 
the Lagrangian $L$ in (\ref{2.1}). 
This will lead to a Hamiltonian system on $\M$ with constraints.
In fact, in the first step we obtain the primary Hamiltonian 
\be
H_P = \frac{1}{2} \langle \J^r, \J^r \rangle 
+ \langle \alpha, \mu - \J^r_{\A^*}\rangle 
+ \langle \beta, \nu - \J^l_{\B^*}\rangle 
+ \langle v_\alpha, \pi_\alpha\rangle  
+ \langle v_\beta, \pi_\beta \rangle
\label{3.9}\ee
together with the primary constraints
\be
\pi_\alpha =0
\quad\hbox{and}\quad
\pi_\beta=0.
\label{3.10}\ee
In addition to being a function on $\M$,
the Hamiltonian $H_P$ contains $v_\alpha\in \A$ and $v_\beta\in \B$, 
which are to be regarded as arbitrary parameters.
We note that $H_P$ is derived from the relation
\be
H_P= p_i \dot{q}^i+ \langle v_\alpha, \pi_\alpha\rangle  
+ \langle v_\beta, \pi_\beta \rangle - L 
\quad\hbox{with}\quad
p_i = \frac{\partial L}{\partial \dot{q}^i},
\label{3.11}\ee
if we restrict to some coordinate neighbourhood $U\subset G$. 
Incidentally, by substituting the explicit formula 
\be
p_i =  \langle \frac{\partial g(q)}{\partial q^i} g^{-1}(q) , \alpha + 
g(q) \beta g^{-1}(q) +\dot g(q) g^{-1}(q)\rangle
\label{p-explicit}\ee 
into the definition (\ref{3.5}),  $\J^r$ and $\J^l$ get converted into 
$J^r$ and $J^l$ as defined in (\ref{2.4}). 
The primary constraints express the fact that $L$ (\ref{2.1}) 
does not depend on the velocities of $\alpha$ and $\beta$.

According to Dirac \cite{Dirac, Sund}, we next  
have to apply a consistency analysis to the system 
$( \M, \{\ ,\ \}, H_P)$ to obtain a constrained manifold
$\M_c \subset \M$ which is preserved by the Hamiltonian 
vector field generated by $H_P$.  
By computing the Poisson brackets $\{ \pi_\alpha, H_P\}$ and
$\{ \pi_\beta, H_P\}$ and noting that these must vanish upon
restriction to $\M_c$, we get the 
secondary  constraints:
\be
\J^r_{\A^*} -\mu =0
\quad\hbox{and}\quad
\J^l_{\B^*}-\nu=0.
\label{3.12}\ee
The derivatives of these constraints  also must vanish along the 
restriction of the Hamiltonian vector field  of $H_P$ to $\M_c$.
It is not difficult to see that this requirement 
leads to the conditions that $\alpha\in \A_\mu$ and 
$\beta\in\B_\nu$, where $\A_\mu$ and
$\B_\nu$ are the Lie algebras of the little groups
$A_\mu$ and $B_\nu$ defined in (\ref{2.2}), respectively.
This means that we must impose the 
further secondary constraints 
\be
\langle  \alpha ,\rho\rangle =0
\quad 
\forall \rho \in \A_\mu^\perp\cap \B^\perp
\quad\hbox{and}\quad
\langle  \beta ,\sigma \rangle =0
\quad \forall \sigma\in 
\B_\mu^\perp\cap \A^\perp.
\label{3.13}\ee 
(For any subspace $W\subset \G$, 
$W^\perp \subset \G$ consists of those $\zeta\in \G$ 
for which $\langle \zeta, w\rangle =0$ holds $\forall w\in W$.)
It is clear that 
these constraints will be preserved by the flow generated by
the Hamiltonian vector field of $H_P$, if we choose 
the so far arbitrary parameters $v_\alpha$ and $v_\beta$ so as to satisfy
\be
v_\alpha \in \A_\mu
\quad\hbox{and}\quad 
v_\beta\in \B_\nu.
\label{3.14}\ee 
The consistency analysis stops at this point.
To summarize, we have arrived at the submanifold 
$\M_c \subset \M$ defined by imposing the constraints 
given by (\ref{3.10}), (\ref{3.12}) and (\ref{3.13}).
The restriction of the Hamiltonian vector field
of $H_P$ to $\M_c$ is tangent to $\M_c$ due to these constraints
together with the restriction (\ref{3.14}).

To continue the Dirac procedure, we have to select the
first class constraints and then find the gauge invariant quantities.
Recall that a constraint $\phi=0$ is first class 
if the Hamiltonian vector field $V_\phi$,
given by $V_\phi[f]=\{ f,\phi\}$ for any  $f\in C^\infty(\M)$,
is tangent to $\M_c$. A function $F$ on $\M_c$ is gauge invariant 
if its derivative is zero with respect to $V_\phi\vert \M_c$ for all
first class constraints $\phi$.
In our case it is not difficult to see that the first class
constraints are 
\be
\langle \pi_\alpha ,\xi \rangle =0,
 \quad
\langle  \pi_\beta ,\eta\rangle =0,
\quad \forall \xi\in \A_\mu,\,\eta\in \B_\nu,
\label{3.15}\ee
and
\be
\langle \J^r_{\A^*}-\mu ,  \xi \rangle =0,
 \quad
\langle \J^l_{\B^*}-\nu , \eta \rangle =0,
\quad \forall \xi\in \A_\mu,\, \eta\in \B_\nu.
\label{3.16}\ee
The momentum constraints (\ref{3.15}) correspond to the gauge 
transformations 
\be
(g, \J^r, \alpha, \pi_\alpha, \beta, \pi_\beta) \mapsto
(g, \J^r, \alpha + \X, \pi_\alpha, \beta + \Y, \pi_\beta) 
\quad\hbox{with some}\quad \X\in \A_\mu, \Y\in \B_\nu,
\label{3.17}\ee
while the gauge transformations generated by the constraints
in (\ref{3.16}) operate as 
\be
(g, \J^r, \alpha, \pi_\alpha, \beta, \pi_\beta) \mapsto
(ag b^{-1}, a \J^r a^{-1}, \alpha, \pi_\alpha, \beta, \pi_\beta) 
\quad\hbox{with some}\quad a\in A_\mu, b\in B_\nu.
\label{3.18}\ee
As a consequence, 
\be
\J^l \mapsto b \J^l b^{-1}.
\label{3.19}\ee 
The translations in (\ref{3.17})  
define an action of the abelian group $\A_\mu \times \B_\nu$ on $\M$, 
where the group structure is given by  the obvious addition, 
and (\ref{3.18}) yields an action of the group $A_\mu \times B_\nu$ on $\M$.
Of course, these gauge transformations map $\M_c$ to $\M_c$.

In analogy to section 2, we now restrict ourselves to
$\check \M \subset \M$ where the factorization 
\be
g= g_A g_B 
\quad\hbox{with}\quad g_A\in A,\, g_B\in B
\label{3.20}\ee
is valid. 
The gauge transformations map $\check \M$ to $\check \M$,
and thus $\check \M_c:= \M_c \cap \check \M$ is also mapped to itself.
On $\check \M$ we have  
\be
(g_A, g_B) \mapsto (a g_A, g_B b^{-1})
\label{3.21}\ee
under the gauge transformations (\ref{3.18}).
It follows that the function $\tilde \L: \check \M \rightarrow \G$ given by
\be
\tilde \L:= g_A^{-1} \J^r g_A = g_B \J^l g_B^{-1}
\label{3.22}\ee
is gauge invariant. 
The formula of $\tilde \L$ can be rewritten as  
\be
\tilde \L= {\tilde \L}_{\A^*} + {\tilde \L}_{\B^*},
\qquad
{\tilde \L}_{\A^*}=\pi_{\B^\perp}(g_A^{-1} \J^r_{\A^*} g_A), \quad
{\tilde \L}_{\B^*}=\pi_{\A^\perp}(g_B \J^l_{\B^*} g_B^{-1}).
\label{3.23}\ee
Defining the function $\L: \check \M_c \rightarrow \G$ 
by $\L:= \tilde \L\vert {\check \M_c}$, we obtain  
\be
\L= \pi_{\B^\perp} ( g_A^{-1} \mu g_A) + \pi_{\A^\perp}(g_B \nu g_B^{-1}).
\label{3.24}\ee 
The components of $\L$ form a complete set among the  
gauge invariant functions on $\check \M_c$. 
In fact, $\L$ parametrizes the space of the gauge orbits in $\check \M_c$,
since this space can be naturally identified with the double coset space
\be
 A_\mu \backslash \check G/ B_\nu \cong A_\mu \backslash A \times B/B_\nu \cong
\O_A(\mu) + \O_B(\nu) \subset \G,
\label{3.25}\ee
where $\O_A(\mu)$ and $\O_B(\nu)$ appear (\ref{1.4}).
We obtain this identification by using that $\alpha$ and $\beta$ 
can be set to zero by the gauge transformations in (\ref{3.17}),
and that on $\check \M_c$ $\J^r$ is uniquely determined by $g$ 
as $\J^r= g_A \L g_A^{-1}$. 

The Dirac brackets of the components of $\L$,
which encode a Poisson structure $\{\ ,\ \}^*$ on the above space
of orbits, can be found by restricting the Poisson brackets of $\tilde \L$
to $\check \M_c$:
\be
\{ \langle \L, \T \rangle, \langle  \L , \T' \rangle \}^* =
\{ \langle  \tilde \L , \T \rangle, 
\langle  \tilde \L , \T' \rangle \}\vert \check \M_c
\qquad \forall \T, \T'\in \G.
\label{3.26}\ee
This relation follows from the standard formula 
of the Dirac bracket \cite{Dirac, Sund}
since $\L= \tilde \L\vert \check \M_c$ and 
$\tilde \L$ have zero Poisson brackets on $\check \M$ with 
all (not only the first class) 
constraints that define $\check \M_c\subset \check \M$.
To calculate the right hand side of (\ref{3.26}), notice that 
\be
\langle  \tilde \L ,\X \rangle = \langle  g_A^{-1} \J^r_{\A^*} g_A , \X\rangle 
\quad\forall \X\in \A,\qquad\quad
\langle  \tilde \L ,\Y \rangle = \langle  g_B \J^l_{\B^*} g_B^{-1} ,\Y\rangle 
\quad\forall \Y\in \B.
\label{3.27}\ee 
By using this, (\ref{3.2}), (\ref{3.4}) and (\ref{3.8}) 
easily lead to the relations
\bea
&&\{\langle \tilde \L, \X\rangle , \langle \tilde \L, \X'\rangle\}=
-\langle \tilde \L, [\X ,  \X']\rangle,
\qquad\forall \X,\X'\in \A,\nonumber\\
&&\{\langle \tilde \L, \Y\rangle , \langle \tilde \L, \Y'\rangle\}=
\langle \tilde \L, [\Y ,  \Y']\rangle,
\qquad\forall \Y, \Y'\in \B,\nonumber\\
&&\{\langle \tilde \L, \X\rangle , \langle \tilde \L, \Y\rangle\}=0.
\label{3.28}\eea  
Thus (\ref{3.26}) implies that the Dirac brackets $\{\ ,\ \}^*$ of the components of
$\L$  are identical to the Poisson brackets $\{\ ,\ \}_*$ (\ref{1.5})
 that appear in the definition of the AKS system.
To identify also the respective Hamiltonians, we note that 
\be
\{ \tilde \L, H_P\} = \{ \tilde \L, \frac{1}{2} 
\langle \tilde \L, \tilde \L \rangle \} 
\qquad
\hbox{on}\qquad \check \M. 
\label{3.29}\ee
Indeed, the last four terms in $H_P$ (\ref{3.9}) have zero Poisson brackets with 
$\tilde \L$ and $\langle \J^r, \J^r\rangle =\langle \tilde \L, \tilde \L \rangle$.
We conclude from (\ref{3.29}) that the Hamiltonian 
\be
H(\L)= \frac{1}{2}\langle \L, \L\rangle
\ee
generates the time evolution of the gauge invariant functions on 
$\check \M_c$ through the Dirac bracket. 
 
In general, the outcome of the Dirac algorithm can be viewed as an effective
Hamiltonian system on a reduced phase space.
The above considerations show that (with the restriction to $\check G\subset G$)
the effective Hamiltonian system 
that belongs to the Lagrangian $L$ in (\ref{2.1}) 
is the AKS system described in the introduction.

We remark that 
if $\check G \subset G$ is a proper submanifold but 
the restriction to $\check G$ is not imposed, 
or the unique factorization appearing in (\ref{2.8}) is 
not valid globally on $A\times B$,
then the application of the Dirac algorithm to the Lagrangian 
(\ref{2.1}) leads to the same Hamiltonian system that 
results also by the corresponding Hamiltonian reduction of
$T^*G$ considered in \cite{RSTS, Reyman, RSTSreview, FT}.

\section{Conclusion}
\setcounter{equation}{0}

The construction described in this paper yields an interpretation 
of certain AKS systems as Lagrangian gauge theories. 
This interpretation is available if the Hamiltonian is the quadratic 
form of a scalar product on a self-dual Lie algebra 
and the factorization in (\ref{2.8}) exists. 

There are many examples (see \cite{RSTSreview})
to which our construction is applicable.
The most familiar case is that of 
$\G= sl(n, \R)$ with $\A$ and  $\B$ being  the strictly upper-triangular 
subalgebra 
and the lower-triangular Borel subalgebra, respectively. 
In this case $\check G$ consists of the Gauss-decomposable elements of
$SL(n,\R)$.
These data can be generalized by 
replacing $sl(n,\R)$ with the normal real form of a simple Lie algebra,
and by using any integral gradation to define a triangular decomposition of $\G$.
Another well known example 
is furnished by taking $\A= so(n,\R)\subset sl(n,\R)=\G$ and 
$\B$ the Borel subalgebra as before.
This example generalizes to any simple Lie algebra, too,
and in the so-obtained cases  $\check G = G$ due to the global nature of the
Iwasawa decomposition.
The open Toda lattices and their various generalizations  
appear among the AKS systems associated with the aforementioned 
Lie algebraic data. Further examples can be found, 
for instance, by using the theory of Drinfeld doubles.

Our definition of the Lagrangian (\ref{2.1}) was motivated by   
the `point particle version' of the 
gauged WZNW model \cite{Balog}
that provides a Lagrangian realization of the Hamiltonian reduction
of the WZNW model to a conformal Toda field theory. 
Since the Lagrangian (\ref{2.1}) is not restricted to Toda systems,
it could be interesting to search for new gauged WZNW
models that would yield field theoretical generalizations 
of the AKS systems treated in this paper.

\section*{Acknowledgments}
This work was supported in part by the Hungarian 
Scientific Research Fund (OTKA) under T034170, T030099, T029802 and M036804.

\end{document}